\documentclass[twocolumn,superscriptaddress,a4paper,nofootinbib,amsmath,amssymb,aps,floatfix]{revtex4-1}
\usepackage{graphicx,color}
\usepackage[normalem]{ulem}
\usepackage{sidecap,dcolumn}

\newcommand{\rt}{\mbox{RFe$_{2}$}}
\newcommand{\rfe}{\mbox{RFe$_{2}$}}

\newcommand{\mssize}{\mbox{$\langle |\mu_{\mathrm{spin}}| \rangle$}}
\newcommand{\mosize}{\mbox{$\langle |\mu_{\mathrm{orb}}| \rangle$}}
\newcommand{\momssize}{\mbox{$\langle |\mu_{\mathrm{orb}}| \rangle / 
                              \langle |\mu_{\mathrm{spin}}| \rangle$}}
\newcommand{\msproj}{\mbox{$\langle \mu_{\mathrm{spin}}^{(z)} \rangle$}}
\newcommand{\moproj}{\mbox{$\langle \mu_{\mathrm{orb}}^{(z)} \rangle$}}
\newcommand{\momsproj}{\mbox{$\langle \mu_{\mathrm{orb}}^{(z)} \rangle / 
                              \langle \mu_{\mathrm{spin}}^{(z)} \rangle$}}
\newcommand{\ms}{\mbox{$\mu_{\mathrm{spin}}$}}
\newcommand{\mo}{\mbox{$\mu_{\mathrm{orb}}$}}
\newcommand{\moms}{\mbox{$\mu_{\mathrm{orb}}/\mu_{\mathrm{spin}}$}}

\newcommand{\tc}{$T_{\text{C}}$}

\newcommand{\ea}{{\it et al.}}
\newcommand{\mm}[1]{\mbox{$#1$}}

\newcommand\unm{%
  \setbox0=\hbox{-}%
  \vcenter{%
    \hrule width\wd0 height \the\fontdimen8\textfont3%
  }%
}

\usepackage{bm}

\begin{document}

\title{Temperature-induced changes in the magnetism of Laves phase
  rare-earth--iron intermetallics by {\em ab~initio}
  calculations}

\author{O.~\v{S}ipr} \email{sipr@fzu.cz}
\homepage{http://crysa.fzu.cz/ondra}
  \affiliation{FZU -- Institute of Physics of the Czech Academy of
  Sciences, Cukrovarnick\'{a}~10, CZ-162~53 Prague, Czechia}
 \affiliation{New Technologies Research
  Centre, University of West Bohemia, CZ-301~00~Pilsen, Czechia}

 \author{S.~Mankovsky}
 \affiliation{Department of Chemistry,
   Ludwig-Maximilians-Universit\"{a}t M\"{u}nchen, Butenandtstr.~11,
   D-81377~M\"{u}nchen, Germany}

\author{J.~Vack\'{a}\v{r}} \affiliation{FZU -- Institute
  of Physics of the Czech Academy of Sciences, Na Slovance~2,
  CZ-182~21 Prague, Czechia}

\author{H.~Ebert}
 \affiliation{Department of Chemistry,
   Ludwig-Maximilians-Universit\"{a}t M\"{u}nchen, Butenandtstr.~11,
   D-81377~M\"{u}nchen, Germany}

\author{A.~Marmodoro} \email{marmodoro@fzu.cz}  
  \affiliation{FZU -- Institute of Physics of the Czech Academy of
  Sciences, Cukrovarnick\'{a}~10, CZ-162~53 Prague, Czechia}

\date{\today}

\begin{abstract}
  Laves RFe$_{2}$ compounds, where R is a rare earth, exhibit
  technologically relevant properties associated with the interplay
  between their lattice geometry and magnetism. We apply {\em
    ab~initio} calculations to explore how magnetic properties of Fe
  in RFe$_{2}$ systems vary with temperature.  We found that the ratio
  between the orbital magnetic moment \mo\ and the spin magnetic
  moment \ms\ increases with increasing temperature for YFe$_2$,
  GdFe$_2$, TbFe$_2$, DyFe$_2$, and HoFe$_2$.  This increase is
  significant and it should be experimentally observable by means of
  x-ray magnetic circular dichroism.  We conjecture that the predicted
  increase of the \moms\ ratio with temperature is linked to the
  reduction of hybridization between same-spin-channel states of atoms
  with fluctuating magnetic moments and to the associated increase of
  their atomic-like character.
\end{abstract}


\maketitle


\section{Introduction \label{sec:introduction}} 

Intermetallic C15-type Laves phase compounds RFe$_{2}$, where R is a
rare earth, exhibit various properties linked to the interplay between
magnetism and the crystal lattice \cite{Stein2004}.  One of them is a
large or unusual magnetostrictive and magnetocaloric effect
\cite{Cla+80,FBI+18,SL+21}. Another reason for interest comes from
thermal or light induced magnetization switching studied intensively
in doped amorphous Gd-Fe alloys with a composition close to GdFe$_2$
\cite{RVS+11,CIK+15,JOL+21}.

An essential part of the research on \rt\ is exploring the various
aspects how electronic states determine magnetism.  An important goal
is understanding the relation between orbital and spin contributions
to the magnetic moments and how these contributions depend on
parameters such as an external magnetic field or temperature.  Recent
interest in the relation between spin and orbital magnetism in
situations when the system is not in the ground state \cite{San+14}
has been motivated, among other reasons, by efforts to understand the
transfer between spin and orbital moments in ultrafast demagnetization
\cite{CMD+08,BBH+10,DSE+21} and by pursuits to comprehend the
temperature-dependence of magnetocrystalline anisotropy \cite{PS+19}.

Concerning the RFe$_{2}$ systems, most attention has been paid to the
magnetic moments of the rare earth component so far.  One of the
reasons is that the experimental research aiming to separate the spin
and orbital moments often employed techniques such as magnetic Compton
scattering combined with vibrating sample or superconducting quantum
interference device magnetometry \cite{IKS+13,MSC+13,AMS+15,YKS+16},
which are sensitive to the sums of rare-earth- and
transition-metal-related contributions.  As the moment of the rare
earth is larger than the moment of Fe, the Fe contribution to
magnetism of RFe$_{2}$ is often hidden behind the contribution of the
R atom.  However, the intriguing properties of the RFe$_{2}$ compounds
are significantly influenced also by the magnetism of the 3$d$
element, because the interatomic interaction between the large 4$f$
moments of the rare earth atoms depends on the exchange coupling
mediated by the transition metal atoms.

Our aim is, therefore, to investigate magnetism of Fe in \rt. For
practical applications as well as for fundamental insight it is
important to understand how magnetic moments change due to finite
temperature.  We employ {\em ab~initio} calculations to investigate
how the spin magnetic moment \ms, the orbital magnetic moment \mo, and
especially the ratio \moms\ vary for \rfe\ if the temperature
increases from zero to the Curie temperature \tc.  To perform thermal
averaging, we employ the alloy analogy model, which proved to be
useful in studying the temperature-dependence of various
magnetism-related properties in the past \cite{EMC+15,SWM+20}.  We
will show below that the \moms\ ratio for Fe in \rfe\ increases with
increasing temperature by as much as 70~\%.


\section{Theoretical scheme} 

\label{sec:methods} 


The calculations were performed within the {\em ab-initio} framework
of spin-density functional theory.  A proper description of magnetism
of rare earths requires special care for exchange and correlation
effects due to the highly localized $f$~electrons.  We used the open
core method in this work: the $f$~electrons were treated as core
electrons, with their numbers fixed during the self-consistency loop
and with the spin and orbital moments set according to the
$LS$~coupling scheme \cite{Rich+01,SME+19}.  Despite its relative
simplicity, many aspects of rare earths magnetism are described
properly within this approach.  Recently, this formalism was employed,
e.g., to get parameters for the Landau-Lifshitz-Gilbert equation to
describe heat-assisted magnetization switching in GdFe$_2$ and
TbFe$_2$ \cite{MKC+19}.  Apart for the open core formalism for the
$f$~electrons, we relied on the local density approximation using the
Vosko, Wilk, and Nusair functional \cite{VWN80}.

\begin{figure}
\includegraphics{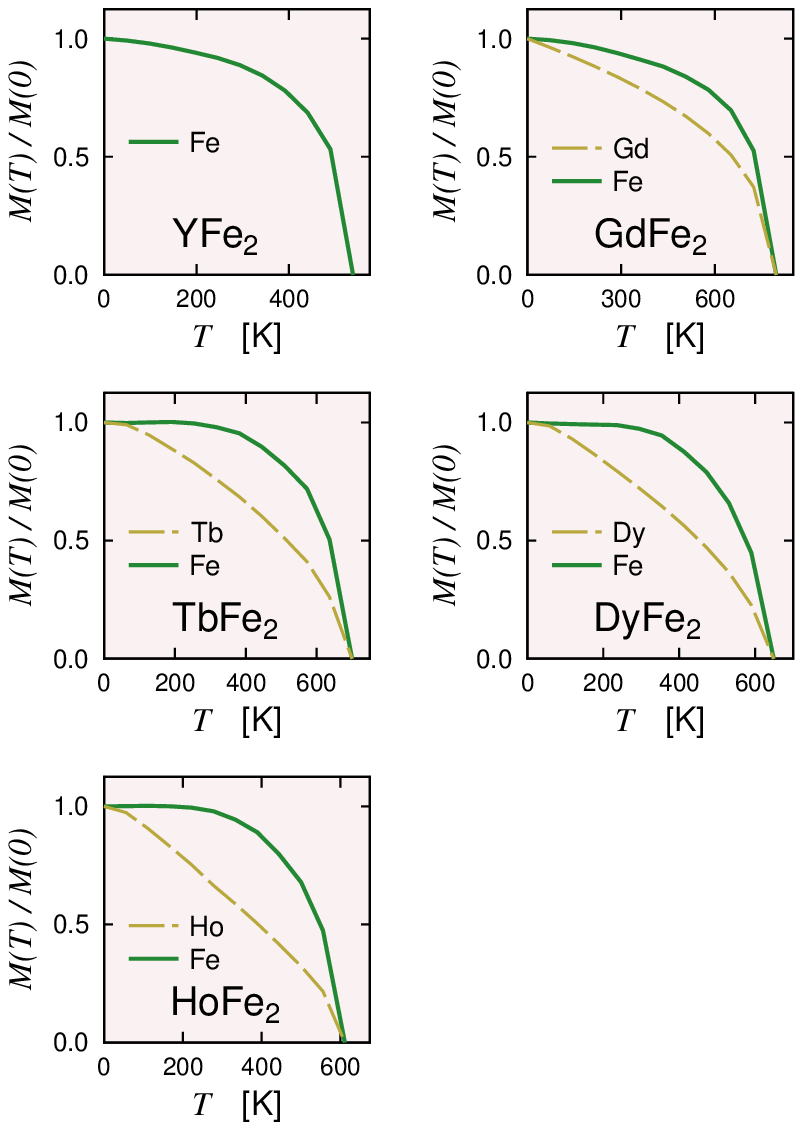}
\caption{\label{fig:mt-curves} Element-specific magnetization curves
  $M(T)$ used as input for our temperature-dependent
  calculations. Data for YFe$_2$ and GdFe$_2$ were taken from Morariu
  \ea\ \cite{MBB+76}, data for TbFe$_2$, DyFe$_2$, and HoFe$_2$ from
  Tang \ea\ \cite{TZL+93}.}
\end{figure}

The temperature effects were included by means of the alloy analogy
model \cite{EMC+15}: temperature-induced atomic displacements and spin
fluctuations are treated as localized and uncorrelated, giving rise to
disorder that can be described using the coherent potential
approximation (CPA).  Atomic vibrations were described using 14
displacement vectors.  Each of the vectors was assigned the same
probability and their amplitude was connected with the temperature by
means of the Debye theory \cite{EMC+15}.  The Debye temperature was
estimated as a weighted average of the Debye temperatures of elements
constituting the RFe$_{2}$ compound.

Spin fluctuations were described by assuming that the local magnetic
moments can get oriented along pre-defined directions; we sampled
60~values for the polar angle and 3~values for the azimuthal angle.
The probability for each orientation was obtained by relying on the
mean-field theory, using as an input the ratio of the magnetization at
the temperature $T$ to the magnetization at $T=0$,
\mm{M(T)/M(0)}\ (see Ref.~\onlinecite{EMC+15} for more details).  Note
that the ratio \mm{M(T)/M(0)}\ is used just to set the probabilities
for pre-defined orientations of the magnetic moments; the values of
respective spin and orbital moments are determined selfconsistently by
the electronic structure calculation itself.  Our approach differs in
this respect from the rigid spin approximation, where it is assumed
that the magnitude of magnetic moments is always the same no matter
what the tilt angle is.

We allowed for atomic-element-dependent magnetization curves $M(T)$
resulting in element-dependent spin orientation probabilities.  The
input data for \mm{M(T)/M(0)}\ used in this work are shown in
Fig.~\ref{fig:mt-curves}.

The calculations were done in a fully relativistic mode using the
spin-polarized multiple-scattering or Korringa-Kohn-Rostoker (KKR)
Green function formalism \cite{EKM11} as implemented in the {\sc
  sprkkr} code \cite{sprkkr-code}.  The potentials were subject to the
atomic sphere approximation (ASA). We verified that this is not a
severe restriction by performing the zero-temperature calculations not
only in the ASA mode but also in the full-potential mode; the
difference in magnetic moments was always less than 5~\%.  The energy
integrals were evaluated by a contour integration on a semicircular
path within the complex energy plane, using a Gaussian-Legendre
quadrature on a mesh of 32~points.  The $\bm{k}$-space integration was
carried out via sampling on a regular mesh, making use of the
symmetry, on a grid of 30$\times$30$\times$30 points in the full
Brillouin zone.  For the multipole expansion of the Green function for
valence electrons, the angular momentum cutoff
\mm{\ell_{\mathrm{max}}}=2 was used.


\section{Results}

\label{sec:results}


\subsection{Temperature-dependence of spin and orbital moments} 

\label{sec:magfe}

\begin{table}
  \caption{Magnetic moments for RFe$_{2}$ (R=Y, Gd, Tb, Dy, Ho)
    calculated for $T=0$~K. The two top lines show moments for the R
    atom, then follow moments for the Fe atoms (for both
    symmetry-inequivalent sites if $\bm{M} \|$[111]), and the last two
    lines show total moments per formula unit.}
\label{tab:moms}
\begin{ruledtabular}
\begin{tabular}{llddddd}
    & & 
  \multicolumn{1}{c}{YFe$_2$} &
  \multicolumn{1}{c}{GdFe$_2$} &
  \multicolumn{1}{c}{TbFe$_2$} &
  \multicolumn{1}{c}{DyFe$_2$} &
  \multicolumn{1}{c}{HoFe$_2$} \\
  &  & 
  \multicolumn{1}{c}{$\bm{M} \|$[111]} &
  \multicolumn{1}{c}{$\bm{M} \|$[111]} &
  \multicolumn{1}{c}{$\bm{M} \|$[111]} &
  \multicolumn{1}{c}{$\bm{M} \|$[001]} &
  \multicolumn{1}{c}{$\bm{M} \|$[001]}  \\
\hline
\ms\ & R     &  0.499  &  7.732 &  6.670 &  5.622 &  4.580 \\
\mo\ & R     & \unm0.003  & \unm0.026 &  2.977 &  4.979 &  5.981 \\[0.5ex]   
\ms\ & Fe-1  & \unm1.863  & \unm2.004 & \unm1.949 & \unm1.919 & \unm1.895 \\  
\ms\ & Fe-2  & \unm1.864  & \unm2.005 & \unm1.951 &
  \multicolumn{1}{c}{\hspace{2.0ex}---} & 
  \multicolumn{1}{c}{\hspace{2.0ex}---}   \\   
\mo\ & Fe-1  & \unm0.044  & \unm0.061 & \unm0.061 & \unm0.055 & \unm0.054 \\[0.2ex]  
\mo\ & Fe-2  & \unm0.053  & \unm0.054 & \unm0.053 & 
  \multicolumn{1}{c}{\hspace{2.0ex}---} & 
  \multicolumn{1}{c}{\hspace{2.0ex}---}   \\[0.5ex] 
\ms\ & tot   & \unm3.228  &  3.722 &  2.769 &  1.784 &  0.790 \\
\mo\ & tot   & \unm0.104  & \unm0.138 &  2.866 &  4.880 &  5.873  
\end{tabular}
\end{ruledtabular}
\end{table}

Magnetic moments for RFe$_2$ (R=Y, Gd, Tb, Dy, Ho) calculated for zero
temperature are presented in Tab.~\ref{tab:moms}.  We show separately
values of \ms\ and \mo\ for the rare earth and for Fe and also the
total moments per formula unit.  The signs are chosen so that the spin
magnetic moment for the rare earth is positive.  The magnetization
direction $\bm{M}$ is set either parallel to [111] or to [001], as
indicated in the header.  This setting was chosen in accordance with
the easy axis of magnetization for each compound
\cite{Tay+71,CAG+78,BSW+03}.  For GdFe$_{2}$ the situation is unclear
\cite{Tay+71,AD+74,MBB+76}, we have chosen $\bm{M}\|$[111].

If the magnetization is parallel to [111], the spin orbit coupling
reduces the symmetry of the system with respect to the case when there
is no spin orbit coupling.  This means that the four Fe sites which
belong to the same type in the absence of spin orbit coupling are
split into two types when it is included: the Fe-1 type comprises one
site and the Fe-2 type comprises the other three sites.  For
$\bm{M}\|$[001] this symmetry reduction does not occur.  The results
presented in Tab.~\ref{tab:moms} are in a good agreement with earlier
theoretical and experimental works for YFe$_2$ \cite{Rit+89,SS+17},
GdFe$_{2}$ \cite{LK+01,SSN+07,MKC+19}, TbFe$_{2}$
\cite{CAG+78,MKC+19}, DyFe$_{2}$ \cite{CAG+78}, and HoFe$_{2}$
\cite{CAG+78,FRS+79}.

\begin{figure*}
\includegraphics{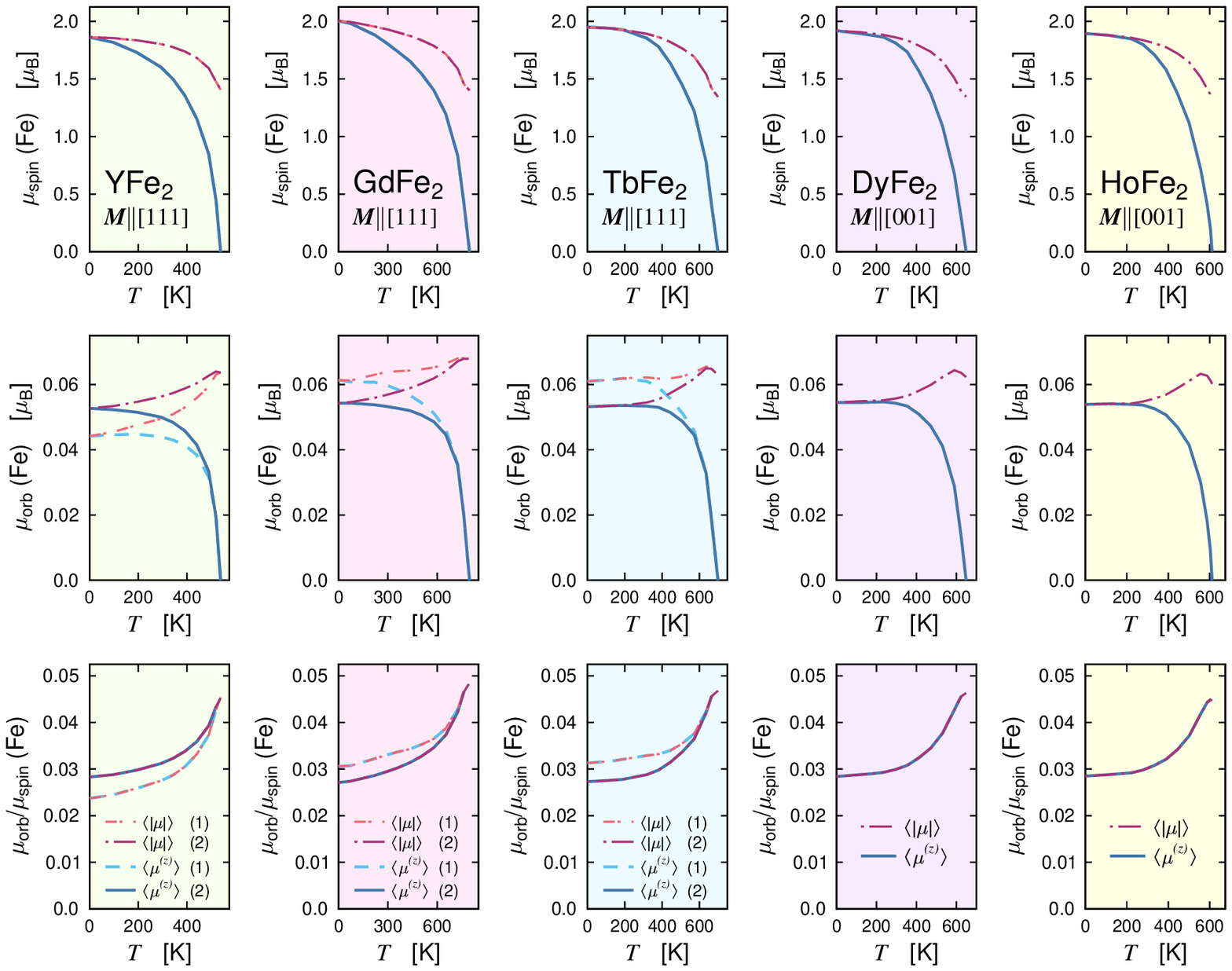}
\caption{\label{fig:mag-t-series} Temperature-dependence of \ms\ (top
  graphs), \mo\ (middle graphs), and \moms\ (bottom graphs) for Fe
  atoms in RFe$_{2}$ (R=Y, Gd, Tb, Dy, Ho).  We present the magnitudes
  of the moments (curves labeled \mm{\langle |\mu| \rangle}) and the
  projections of the moments on the common magnetization direction
  (curves labeled \mm{\langle \mu^{(z)} \rangle}). The numbers (1) or
  (2) appearing in the legend for $\bm{M}\|$[111] distinguish between
  symmetry-inequivalent sites.  }
\end{figure*}

The calculated temperature-dependence of \ms, \mo, and \moms\ for the
Fe sites is shown in Fig.~\ref{fig:mag-t-series}.  We monitor two
quantities, namely, (i) the magnitude of the magnetic moment and (ii)
its projection on the easy axis of magnetization.  These quantities
were obtained by evaluating their configuration average by means of
the alloy analogy model (Sec.~\ref{sec:methods}).  We choose the
notation \mm{\langle |\mu| \rangle}\ for the magnitude of the moment
and \mm{\langle \mu^{(z)} \rangle}\ for its projection, with $\mu$
representing the magnetic moment (spin or orbital).  By the
superscript $(z)$ we always mean the common magnetization direction
(in our case, either [111] or [001]).

Concerning the projections of the magnetic moments, one can observe by
inspecting Fig.~\ref{fig:mag-t-series} that if the temperature
increases, both \msproj\ and \moproj\ decrease so that they are zero
at $T=T_{C}$, as mandated by the model. However, the situation differs
for the magnitudes of the magnetic moments, where \mssize\ still
decreases with increasing $T$ but to a finite nonzero value at
$T=T_{C}$, whereas \mosize\ {\em increases} with $T$.  The fact that
the local magnetic moments for Fe atoms are nonzero even at $T_{C}$
resemble the behavior of Fe in other systems \cite{RKM+07}.  For the
ratio \moms\ it does not matter whether we evaluate the ratio of the
projections \momsproj\ or the ratio of the magnitudes \momssize, both
quantities increase with increasing $T$ in the same way.  This
increase is by about 70~\% in the interval from $T=0$~K to $T=T_{C}$.

As mentioned in connection with Tab.~\ref{tab:moms}, there are two
inequivalent Fe atomic types for $\bm{M}\|$[111].  This distinction
does not lead to apparent differences concerning the values of
\ms\ but there are clear differences for \mo.  If the temperature
increases, these differences decrease (cf.\ the middle graphs in
Fig.~\ref{fig:mag-t-series}).

Further analysis shows that the temperature dependence of the magnetic
moments is associated mainly with the fluctuations of the spin moments
directions.  Namely, additional calculations demonstrate that if the
lattice vibrations are ignored, the results presented in
Figs.~\ref{fig:mag-t-series} practically do not change (results not
shown).

The electronic structure of the rare earth elements (Gd, Tb, Dy, Ho)
is treated within the open core approach.  The magnitude of the
corresponding magnetic moments is thus temperature-independent and
their projection on the common magnetization axis $\bm{M}$ is just the
average of the projections of the pre-defined spin directions weighted
by corresponding thermodynamic probabilities (see
Sec.~\ref{sec:methods}).  In other words, the temperature dependence
of the $f$~electrons-related moments simply reproduces the respective
input $M(T)$ curves shown in Fig.~\ref{fig:mt-curves}.  As the
$f$-electrons moments are dominant in rare earths, the same applies
also to total spin and orbital moments of Gd, Tb, Dy, and Ho atoms.
For this reason we do not show the curves here.  Magnetism of Y in
YFe$_2$ is induced by Fe and its temperature-dependence (not shown)
therefore completely follows the temperature-dependence of the Fe
moments.


\subsection{Influence of a moment tilt on \moms} 

\label{sec:tilt}

\begin{figure}
\includegraphics{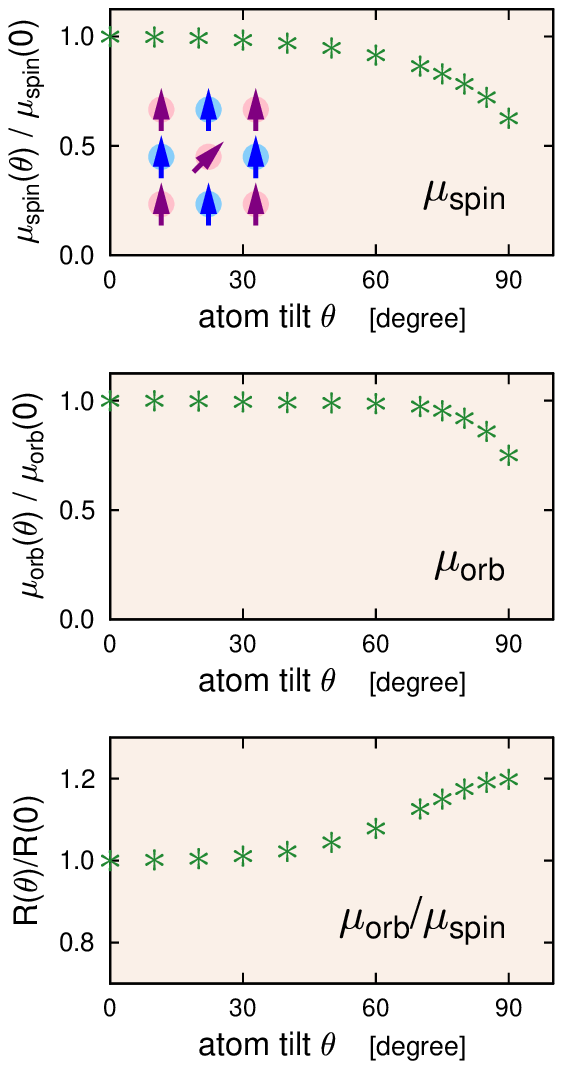}
\caption{\label{fig:tilt} Dependence of \ms, \mo, and the ratio
  $R$=\moms\ for an Fe atom in GdFe$_{2}$ on the angle~$\theta$ by
  which the moment for a single Fe atom is rotated. The quantities were
  evaluated for $T=0$~K and normalized to their values at
  $\theta=0^{\circ}$.}
\end{figure}

For further understanding, we examine what happens if the
moment at a single Fe atom is gradually tilted with respect to the
rest, taking GdFe$_{2}$ as an example.  We assume that all magnetic
moments are oriented along the same direction except for a single Fe
atom, where the moment is tilted by the angle~$\theta$.  The results for
the magnitudes of spin and orbital moments and for their ratios are
summarized in Fig.~\ref{fig:tilt}.  One can see that the magnetic
moment of a tilted Fe atom in GdFe$_{2}$ remains finite even for
$\theta=90^{\circ}$.

This provides another view on the \moms\ ratio for the Fe atoms.
Namely, the bottom graph of Fig.~\ref{fig:tilt} shows that
\moms\ increases for Fe atoms with an increasing magnetization tilt
angle.  This is consistent with the increase of the \moms\ ratio with
increasing temperature demonstrated in Fig~\ref{fig:mag-t-series}:
within the alloy analogy model, larger temperature means bigger
statistical weight of configurations with larger tilt angle (see
Sec.~\ref{sec:methods}).


\section{Discussion}

\label{sec:discuss}

We predict for several \rfe\ compounds that the ratio \moms\ increases
with increasing temperature.  This prediction should be verifiable by
x-ray magnetic circular dichroism (XMCD) experiments, because even
though evaluating \ms\ and \mo\ separately by means of the XMCD sum
rules \cite{CTAW93,TCSvdL92} is not always straightforward, the
determination of their ratio \moms\ is much more robust and can serve
as an indicator of even quite subtle effects \cite{SMS+11}.  In
addition it should be noted that XMCD was already used for studying
the temperature dependence of \moms\ for Sm atoms in Laves compound
Sm$_{0.974}$Gd$_{0.026}$Al$_2$ \cite{DLB+10}.  Analyzing corresponding
data for Fe in RFe$_2$ systems would be more difficult because the
magnetic moment of Fe is smaller than the magnetic moment of rare
earth atoms and it further decreases for elevated temperatures where
the increase of \moms\ is to be observed.  However, this should not be
a severe problem.  Our data indicate that for temperatures for which
\moms\ increases by 50~\% with respect to $T=0$~K, the magnetic moment
for the Fe atoms decreases just to about half of its zero-temperature
value (see Figs.~\ref{fig:mt-curves}--\ref{fig:mag-t-series}).  In
such situation, an Fe $L_{2,3}$-edge XMCD signal should be strong
enough to be clearly resolved.

An intuitive explanation for the increase of \moms\ with increasing
temperature might be sought based on the lower graph of
Fig.~\ref{fig:tilt}.  Here we see that if the magnetic moment of a
single Fe atom is tilted, the ratio \moms\ increases; for the tilt
angle $\theta=90^{\circ}$ this increase is 20~\%.  Tilting the moment
means reducing the hybridization between the states of the tilted atom
and the states of surrounding atoms belonging to the same spin
channel.  In other words, the electron states associated with the
tilted Fe atom become more atomic-like.  In this limit, the experience
shows that the \moms\ ratio generally increases with respect to the
bulk --- compare, e.g., with the situation for clusters or surfaces
\cite{SKE+04,SKZ+10}.  Increasing the temperature means increasing the
statistical weights of atoms with tilted moments.  We thus conjecture
that the increase of \moms\ with increasing temperature is linked to
the reduction of the hybridization within the same spin channel
between states of tilted Fe atoms.  In this regard, one can see an
analogy with the enhancement of \mo\ at surfaces: in that case the
hybridization between the states of surface atoms is smaller than the
hybridization between the states of bulk atoms.  Analogously in our
case, the hybridization between the states of atoms with tilted
moments is smaller than the hybridization between the states of atoms
with parallel moments.

The increase of \moms\ with increasing temperature appears to be quite
a robust effect for RFe$_2$ systems.  It occurs for any orientation of
the magnetization and for any choice of R we considered.  Note that R
does not need to be a rare earth element --- the effect occurs also
for YFe$_2$.  One can expect that a similar effect, namely,
temperature-related increase of the \moms\ ratio, should occur for
other systems as well.  Prime candidates would be compounds with Fe
atoms which have non-zero magnitude of the local magnetic moment even
at temperatures close to \tc.

Let us now turn to the difference in \mo\ for symmetry-inequivalent Fe
sites, as it happens for $\bm{M}\|$[111] (see the graphs for YFe$_2$,
GdFe$_2$, and TbFe$_2$ in Fig.~\ref{fig:mag-t-series}).  If the
temperature approaches \tc, the difference between \mo\ at both sites
gradually disappears.  This is a consequence of the increased disorder
in the spin directions when the temperature increases.  At $T=T_{C}$
this disorder gets maximum, the magnetization direction $\bm{M}$
ceases to have any meaning and the Fe sites become equivalent,
similarly as they would be in the absence of spin-orbit coupling.


\section{Conclusions}

\label{sec:conclusions}

Ab-initio calculations with finite-temperature effects included by
means of the alloy analogy model predict that the \moms\ ratio
increases with increasing temperature for \rfe\ Laves compounds. We
interpret this as a consequence of reduced hybridization between
states (within the same spin channel) of atoms with fluctuating spin
magnetic moments and the associated increase of their atomic-like
character if the temperature increases.  The predicted enhancement of
\moms\ is large enough to be seen by XMCD experiments.


\begin{acknowledgments}
  This work was supported by the GA~\v{C}R via the project 20-18725S.
  Additionally, computing resources were supported by the project
  CEDAMNF CZ.02.1.01/0.0/0.0/15\_003/0000358, by the project
  ``e-Infrastruktura CZ'' (e-INFRA CZ LM2018140) and by the project
  ``Information Technology for Innovation'' (IT4I OPEN-22-42) provided
  by the Ministry of Education, Youth and Sport of the Czech Republic.
\end{acknowledgments}


%

\end{document}